\documentstyle[aps,preprint,eqsecnum]{revtex} 
\input{epsf}
\begin{document} 
\title{On the foundation of equilibrium quantum statistical mechanics}
\author{Giulio Casati} 
\address{International Center for the Study of Dynamical Systems,
Via Lucini 3, 22100 Como, Italy\\
Istituto Nazionale di Fisica della Materia and INFN, sezione di Milano,
Italy}
\draft

\maketitle

\begin{abstract} 

We discuss the condition for the validity of equilibrium quantum statistical
mechanics in the light of recent developments in the
understanding of classical and quantum chaotic motion.  In
particular, the ergodicity parameter introduced in \cite{CaCh} is
shown to provide the conditions under which quantum statistical 
distributions can be derived from the quantum dynamics of a classical
ergodic Hamiltonian  system.

                    D Y S C O  80

                     October 1997
\end{abstract}
\pacs{
\hspace{1.9cm} 
PACS number 05.30.-d}

\section{\bf Introduction}

The pioneering works of Fermi, Pasta, Ulam(FPU) \cite{fpu} and the
mathematical results originating from the ideas of Kolmogorov  (see, 
e.g., Refs.\cite{KOL}) in the mid  fifties, have started a new
era in our understanding of the behaviour of dynamical systems.
At their first appearance these results were considered by most
physicists as merely an interesting curiosity with no real physical
relevance.
The dominant belief was that, though they were correct, the large number
of
degrees of freedom and the complexity of the interaction of physical
systems would inevitably lead to the validity of traditional statistical laws.
For example these results did not shake the great confidence in the
postulate of ``{\it a priori\/} equiprobability" or the Boltzmann ergodic
hypothesis.  It was generally believed that there is nothing in the
laws of
mechanics which would lead to expect that an isolated system is
more likely to be in one of its accessible states than in
another. The validity of this view was supported by the
agreement of theoretical predictions with experimental
observations. 

More recently, examples of dynamical systems have been found, for
which the ergodic hypothesis has been rigorously shown to be
valid\cite{Sin}. In particular it has become clear that for the
validity of statistical laws the so-called thermodynamic limit
is not necessary. The property of mixing or positive KS
entropy is sufficient to ensure good statistical behaviour provided 
the  number
of degrees of freedom $N>2$. Thus, an isolated
system, no matter what the initial condition may be, will indeed reach a
final equilibrium situation in which it is equally likely to be
found in any one of its accessible states.
In terms of distribution functions one may consider an ensemble
of systems which are initially in some subset of the accessible
states. During  time evolution these systems will make transitions between the various
accessible states until they are uniformly distributed and this will
correspond to the final equilibrium situation. In this connection we would like to recall that the process of statistical
relaxation is obviously time-reversible but the evolution of the
distribution function is non recurrent.

Once the above mixing property has been assumed or shown to hold, then statistical averages of physical quantities can be
performed via the microcanonical or the equivalent canonical
ensemble.  To be more specific, let us consider a conservative, 
Hamiltonian system with N degrees of freedom:
\begin {equation}
H\, =\, H_0\, +\, V
\label {eqclH}
\end {equation}
where $H_0$ is some integrable Hamiltonian, for example a system
of N independent harmonic oscillators, and $V$ a  non
integrable perturbation. If the perturbation $V$ is such that it
renders the hamiltonian (\ref{eqclH}) ergodic and mixing, then, 
for any initial state, the system will approach a
microcanonical equilibrium. As a consequence, the equipartition
theorem can be rigorously proven, which means that, for example, 
for sufficiently small $V$ there is energy equipartition among
the different oscillators. Analogously, since the
microcanonical ensemble implies equal probability for equal
regions of the energy surface, it follows that 
the most probable distribution of
particles among their own individual states for a system in
macroscopic equilibrium is given by the Maxwell Boltzmann law\cite{tolm};
 namely, the probability that a particle
has energy $\epsilon$ is proportional to $exp( -\alpha\epsilon)$.

On the other hand it is now known that, if the perturbation is
small enough then, generically, invariant regions of positive measure exist on the
energy surface, most orbits lie on  N-dimensional tori and
typically the motion is ergodic over these tori only (and not on the
2N--1 dimensional energy surface). Therefore, the microcanonical
ensemble cannot be used to compute statistical averages and the
equipartition theorem does not hold: different regions on the
energy surface are not equally probable and the average
equilibrium values strongly depend on the initial conditions.

Certainly, the problem remains open to establish, for any given
system, whether it is ergodic or not, with positive KS entropy
or not etc. In other words, it is not
pratically known, rigorously speaking, when one can in fact be
sure of the validity of the microcanonical ensemble. (We know from the
 Siegel theorem\cite{siegel} that in
the space of Hamiltonian systems the vast majority of them is
non integrable.) Actually, an entire new field, 
chaos and dynamical systems, has arisen from this problem.
Nevertheless, at present, as far as the foundation of statistical mechanics is concerned, 
 the situation is in principle quite clear.

\section{\bf The quantum ergodic problem}

The problem now is what happens in quantum mechanics, which is
the subject of the present paper. Existing standard textbooks
assume the validity of the microcanonical (and canonical) ensemble
and then proceed to derive, in several different ways, 
Bose-Einstein or  Fermi-Dirac statistics, depending on the
symmetry requirements.The idea behind this is to devise a representative
ensemble of systems and then take the average properties of
the systems in this ensemble as good estimates for quantities
pertaining to the system of actual interest. The representative
ensemble is constructed according to the postulate of  {\it a priori\/}
pobability which is considered as a {\it non arbitrary} postulate.
According to existing standard textbooks on the subject\cite{tolm} 
``... it is also evident that it would be arbitrary to proceed
otherwise than by assignment  of random phases, since the
quantum mechanics has not itself provided any reason for
thinking that any particular arrangement of phases is inherently
more probable than a random one... We now see that our
postulate, of equal {\it a priori\/} probabilities and random {\it
a priori\/} phases for different quantum mechanical states, is
equivalent at the correspondence principle limit to the
assumption of equal {\it a priori\/} probablities for different
regions of equal extension $h^f$ in the classical phase space ...
the postulate actually introduced is the only
non arbitrary one that can be selected, and agrees at the
correspondence principle limit with that selected for the
classical statistic. The methods developed do have, as far as is
known, the {\it a posteriori} justification of agreement with
experimental findings". Quite clearly, this point of view has no solid theroretical grounds 
and indeed, as remarked in \cite {kh},  ``... such a point of view is not
entirely
satisfactory because these postulates cannot be independent of, and should
be derivable from, the quantum mechanics of molecular systems. A rigorous
derivation is at present lacking"

At a more rigorous level, the problem of quantum ergodicity was 
discussed for the first time in a famous paper by Von Neumann\cite{neu}. 
In this paper Von Neumann established an inequality which he thought gave  
a dynamical foundation to quantum ergodic theory. In a series of
papers \cite{paul,ter,far,prosp} the results of Von Neumann were 
strongly criticized and in particular in \cite{prosp} it was shown that
the Von Neumann inequality was entirely a consequence of the averaging
over ``macro-observers" and has nothing to do with quantum dynamics. It is
not our purpose here to discuss why all previous
numerous attempts\cite{neu,paul,ter,far,prosp,janc,farq} to prove
the quantum
ergodic theorem failed. It was not even clear how
to formulate the problem, and different ways had been proposed
to circumvent the difficulty in formulating a quantum parallel
of the classical ergodic theorem; namely,  the fact that when the
system is in an energy eigenstate, the probabilities do not
change in time. For an extensive discussion we refer to \cite{janc,farq} 
and to the more recent review by Pechukas \cite{pec}. In conclusion, the 
problem was left open: the main point is
that  all previous attempts did not call into  question the structure
of the eigenfunctions themselves, which is instead the crucial
issue. In other words, in the computations of time-averages, the
structure of the hamiltonian and therefore the dynamics, did
not enter!

We now have a much better understanding
of the dynamical properties of both classical and quantum systems.
However, a critical re-examination of the fundamental
hypothesis, as has been made in classical mechanics after the
work of FPU, has not yet been made in quantum mechanics. This is
quite surprising, expecially in consideration of the increasing
interest in the so-called ``quantum chaos", which deals mainly
with the properties of quantum systems which are classically
chaotic.  In this respect a large number of papers is devoted to
correlation properties of levels, periodic orbit theory and the
like, but, as far as we know, a critical reexamination of the 
foundations of quantum statistical mechanics has not yet been 
undertaken\cite{nota}. For example, it is interesting that, while the
justification of the
microcanonical ensemble in classical mechanics has been the
object of lively debates for almost one century, in
quantum mechanics, as we have seen, the microcanonical ensemble
is simply assumed and the only justification remains a hand-waving 
analogy with classical ergodicity and the agreement with
experiments. 

First of all, it is now clear that the existence of  N-dimensional 
invariant surfaces in classical systems implies restrictions on the
quantum wavefunctions. Certainly, while
an invariant KAM(after Kolmogorov, Arnold, Moser) curve constitutes an
impenetrable barrier for
the classical orbit, in quantum mechanics there is tunneling
outside the invariant curves; however this tunneling is
typically exponentially small and the situation is quite similar
to the classical one. Therefore, the different complex problems
generated by the presence of a divided phase space in classical
mechanics, must also be taken into account in quantum mechanics. In
particular, the lack of equipartition in classical mechanics implies a corresponding deviation from the expected equilibrium quantum statistical distributions. That is, the
equilibrium properties will depend, as in classical mechanics, 
on the initial state.  

However, our main interest here is in systems which are classically ergodic.
In the following we will show that, besides classical ergodicity, 
additional
conditions must be satisfied by the quantum Hamiltonian for the validity of
equilibrium quantum statistical distributions.  As a matter of fact, there 
is no reason to
believe that the ergodicity of a classical system can be taken
as  justification for the assumption of  {\it
a priori equiprobability\/} in quantum mechanics. On the contrary, 
the so-called {\it quantum dynamical localization}, which is one of the
main modifications imposed by quantum mechanics on  classical chaotic
motion, seems to point in the opposite direction. This phenomenon was
discovered 20 years ago in the model of the kicked
rotator\cite{CaCh,bible} and has now been observed in several laboratory
experiments\cite{ba,ko,Wa,Ra}. It consists in the
suppression of the diffusion process generated by  classical
chaotic motion and parallels the well- known Anderson
localization. The crucial difference is that, in contrast to
Anderson localization, it takes place in systems without any
disorder in the Hamiltonian. More recently, dynamical
localization has also been shown to take place  in conservative
systems\cite{CCGI,YaF,shell,borg}. For recent reviews see
also\cite{CaCh,CaChCo}. It is our purpose to understand how
dynamical localization manifests itself in systems of type
(\ref{eqclH}), classically ergodic, and what is the relation of this new recently
discovered phenomenon with the problem at hand.

\section{\bf Localization in quantum Hamiltonian systems}

In order to approach this problem conveniently, let us analyze a
model of conservative systems  introduced by
Wigner \cite {W} and known as the Wigner Band Random Matrix (WBRM) 
model.  Namely, we consider an ensemble of real hamiltonian
matrices of a rather general type:
\begin{equation} 
 H_{mn}\, =\epsilon_n\, \delta_{mn}\, +\, v_{mn}\qquad(m, n=1,.., N) 
\label{eqWm} 
\end{equation} 
where the off--diagonal matrix elements $v_{mn}=v_{nm}$ are
statistically independent, Gaussian random variables, with
$<v_{mn}>=0$ and $<v_{mn}^2>=v^2$, if $|m-n|\leq b$, and are
zero otherwise.  The WBRM  (\ref{eqWm}) provides a good
description of the quantum statistical properties of general Hamiltonian 
systems of type (\ref{eqclH}) and therefore their consideration
is very appropriate for the purposes of our discussion.
The matrix (\ref{eqWm}) 
is given in the basis of unperturbed eigenstates $\phi_n$ of
${\hat H}_0$. Although in
completely integrable quantum systems there is a quantum number
for each degree of freedom, we suppose that the unperturbed
states are ordered according to increasing energy, and we
thereby label them with a single number $n$.  We denote by $\rho$
the average density of states
\begin{equation} 
 \rho^{-1}\, =\, \left<\epsilon_n-\epsilon_{n-1}\right> 
\label{eqdl} 
\end{equation} 
where the averaging is understood either over disorder or within
a single, sufficiently large, matrix. Both ways are equivalent
owing to the assumed independence of matrix elements.

Let us consider  the EF matrix $C_{mn}$, which connects exact
eigenfunctions $\psi_m$, obtained by diagonalization of the
Hamiltonian matrix (\ref{eqWm}), to unperturbed basis states
$\phi_n$, 
\begin{equation} 
 \psi_m\, =\, \sum_n\, C_{mn}\cdot\phi_n  
\label{eqefm} 
\end{equation} 
{}From the matrix $C_{mn}$ we can compute both the statistical distribution  
$W_m(n)=C^2_{mn}$ of the eigenstates $\psi_m$ on the unperturbed ones 
$\phi_n$, and the distribution $w_n(m)$ of the unperturbed  
eigenstates on the exact ones. These distributions have 
classical analogs $W$ and $w$. Indeed, in the classical case, the
unperturbed energy $E_0$ is not constant along a classical
chaotic trajectory of the full Hamiltonian with a given total
energy $H=E$. Instead, it sweeps a range of values, or ``energy
shell", $\Delta E_0=\Delta V$, and is distributed inside this
shell according to a measure  $W_{E}(E_0)$. The form of
$W_E(E_0)$ depends on the form of the perturbation $V$; we will
call this measure ``ergodic" because it is determined by the
ergodic (microcanonical) measure on the given energy surface
$H=E$.  The quantum analog of this measure characterizes the
distribution of the ``ergodic" eigenfunction  in the
unperturbed basis.  Conversely, if we keep the unperturbed
energy $E_0$ fixed,  the bundle of trajectories of the total
Hamiltonian $H$, which reach the surface $H_0=E_0$, has a
distribution in the total energy $E$ which is described by a
measure $w_{E_0}(E)$.  In the quantum case, this measure
corresponds to the energy spectrum of the Green's function 
at energy $E_0$. It is also called the local spectral density of states (LDOS) 
$w_{E_0}(E)$ or strength function or spectral measure of the
unperturbed eigenstate at energy $E_0$. For a typical
perturbation, represented by a WBRM, the average $w(E)=\langle
w_{E_0}(E)\rangle$ depends on the Wigner parameter

\begin{equation} 
 q\, =\, \frac{(\rho \, v)^2}{b} 
\label{eqq} 
\end{equation} 
and has the following limiting forms 
 \cite{W} (see also Refs.\cite{CaChCo,YaF}) 
\begin{equation} 
 w(E) \, =\, \left\{  
\begin{array}{ccc} 
 \frac{2}{\pi E_{sc}^2}\, \sqrt{E_{sc}^2\, -\, E^2}, &  
|E|\, \le\, E_{sc}, & q\, \gg\, 1\\ 
 & & \\
 \frac{\Gamma /2\pi}{E^2\, +\, \Gamma^2/4}\cdot 
 \frac{\pi}{2\cdot\arctan{(1/\pi q)}}, & |E|\, \le\, E_{BW}, & q\, \ll\, 1 
\end{array} \right. 
\label{eqBW} 
\end{equation} 
Outside the specified energy intervals, both distributions have
exponentially small tails.  In the limit $q>>1$ we have
the semicircle law with a width of the energy shell $\Delta E =
2E_{sc}=4v\sqrt{2\, b}$. In the other limit, 
$q<<1$, we have the Breit - Wigner  distribution, of width
$\Delta E = 2E_{BW}=2b/\rho$  with the main part of the distribution inside
 a width $\Gamma = 2\pi\rho v^2 $.  In all these
expressions $E$ is measured with respect to the center of the
distribution.

The phenomenon of localization is related to the possibility
that the eigenfunctions are localized on a scale which is 
significantly smaller than the maximum one consistent with energy
conservation.  Indeed, the localization length, namely the size of the
region which is populated
by an eigenfunction,  is bounded from above by the {\it ergodic
localization length\/} $d^{(e)}=c\rho\Delta E$, which measures the
maximum number of basis states coupled by the perturbation.
This length characterizes the full width of the energy shell
$\Delta E$. (The factor $c$ depends on the particular definition of
localization width). In other
words, in a conservative quantum system there is always
localization in energy, due to the existence of a finite $\Delta
E$ \cite{CCGI}. This fact, which is sometimes a source of
confusion,  is  just  a trivial consequence of energy
conservation. What is physically relevant is the possibility to have localization {\it
inside\/} the shell \cite{CCGI}. This possibility depends on a scaling
parameter $\lambda$ which we call the ``ergodicity parameter". Indeed, in
\cite{CCGI} it was shown that  the average  
localization length $d$ of eigenfunctions defined as $d\equiv\langle (\sum_n
W^2_m(n))^{-1}\rangle$ (inverse participation ratio) obeys a scaling law of the form
\begin{equation}
\beta_d\, =\, \frac{d}{d^{(e)}}\, \approx\, 1\, -\, {\rm
e}^{-\, \lambda}
\label{eqdde} 
\end{equation} 
where 
\begin{equation} 
\lambda\, =\, \frac{ab^2}{d^{(e)}}\, =\, \frac{ab^{3/2}}{\rho v} 
\label{erpar} 
\end{equation} 

Here $a\approx 0.23$ \cite{shell}.  The parameter $\lambda$  
plays the role of an {\it ergodicity parameter\/} because, 
when it is large, the localization length approaches its maximal
value $d^{(e)}$, which means that the eigenfunctions become
ergodic, i.e., delocalized over the whole energy shell. Instead, if
$\lambda<<1$, 
the eigenfunctions are strongly localized inside the energy shell.
Notice that the matrix size $N$ is an
irrelevant parameter, provided $N\gg d^{(e)}$ is large enough to
avoid boundary effects.

In fig(1) we show a typical example of a localized
eigenfunction(solid line). This figure is obtained\cite{shell} from a single
matrix with parameters $N=2400$, $v=0.1$, $b=10$, $\rho=300$. In order to
suppress fluctuations in individual distributions, averages have
been taken over 300 of them, chosen around the center of the
spectrum. The solid line is obtained by averaging with respect to
the center of each eigenfunction so that the typical structure
of the eigenfunction is revealed. The center is defined as:
\begin{equation}
n_c(m) = \sum_n W_m(n) n
\end{equation}
The circles are obtained instead by averaging the same
eigenfunction with respect to the center of the energy shell
namely counting the site label in $W_m(n)$ starting from the
reference site $m$ (instead of $n_c(m)$). As discussed in
\cite{shell}, this second type of average is expected to agree with the
LDOS. 

As  is seen in fig. (1),
the actual width of the eigenfunction is much less than the width of
the  LDOS
which gives the maximum number of unperturbed states that
can be coupled by the perturbation. This means that the eigenfunction 
is not ergodic.

Even though the analysis presented in this section is based on a
random matrix model 
of a conservative system, the results remain qualitatively the same for
 real quantum conservative systems. Indeed it is now 
well established that the mechanism of localization inside the energy shell
is quite typical
and it was recently shown to take place e.g.in a classically chaotic 
billiard\cite{borg}.

\section{\bf Quantum statistical distributions}

In the previous section we have shown that occurrence of dynamical
localization in Hamiltonian systems leads to a deviation from ergodicity.
 A similar non ergodic quantum behaviour takes place when
quantizing  non ergodic classical systems. Indeed, 
if  there are invariant regions of positive
measure on the energy surface, such as invariant tori, then the range
of values of $E_0 $ swept by a classical orbit will be much 
narrower, or ``localized" inside the classical energy shell and 
will moreover  depend on initial conditions. In such a 
situation for example, the classical Maxwell Boltzmann distribution will not follow. For exactly the same reason, 
the equilibrium quantum statistical distribution is not to be expected
in the presence of quantum localization independently of whether the
latter is produced by purely dynamical quantum effects or by the
presence of islands of stability in the classical phase space. 
In conclusion, even if the classical system is ergodic and mixing, 
the phenomenon of quantum localization
{\it do indeed provide reasons\/} to reject the
notion that the equal {\it a priori\/} probability is a non arbitrary
postulate which agree with the correspondence principle. The quantum steady state will depend on the initial conditions and only in the classical limit will approach the microcanonical distribution in accordance with the correspondence principle. 

In order to explicitely show  the mechanism through which quantum
localization prevents the derivation of  quantum statistical
distributions, let us  consider one of the standard methods
 \cite{reif} (see also \cite{ff}) to derive equilibrium distributions:
 
The average occupation number, $n_s$, is given by
\begin {equation}
\overline{n}_S(E) = {\sum_\ell n_S(\ell) W_E
(E_\ell) \over { \sum_\ell W_E
(E_\ell) }}
\end{equation}

In order to compute the above average, according to  \cite{reif}, one 
introduces the sum
extended over all states except  state $s$: $Z^s_{E} (N) = \sum^{(S)}_\ell W_E
(E_\ell)$.                     
Then, in  Fermi Dirac case $(n_s = 0, 1)$ for example, one obtains
\begin{equation}
\overline{n}_S(E) = { 0 + Z^s_{E-\epsilon_S} (N-1) \over { 
Z^s_{E-\epsilon_S} (N-1) + Z^s_{E} (N) }} = 
{1 \over {1 + Z^s_{E} (N) / Z^s_{E-\epsilon_S} (N-1) }}
\end{equation}
Since the numbers of terms  in the sum for the partition function is
large, one can write
\begin{equation}
ln Z^s_{E-\epsilon_S} (N-1) = ln Z^s_{E} (N) -\alpha_S - \beta_S
\epsilon_S
\end{equation}
where $\alpha_S = {\partial ln Z^s_{E} (N) \over { \partial N}}
$, $\beta_S = {\partial ln Z^s_{E} (N) \over { \partial E}}$.
This leads to
\begin{equation}
\overline{n}_S(E) = {1 \over {1 + e^{\alpha_S+\beta_S \epsilon_S}}}.
\label{fd}
\end{equation}

Now, since the sum defining $ Z^s_{E} (N)$  is over many states, one expects that the
parameters $\alpha_S$  and $\beta_S$  are insensitive as to which particular
state $s$ is omitted from the sum and therefore one can assume
\begin{equation}
\alpha_S =\alpha ;
\beta_S=\beta={1 \over T}.
\end{equation}
Then (\ref{fd}) gives the well known Fermi Dirac distribution.

However, in case of quantum localization, not only are the
eigenfunctions  localized inside the shell, but, as clearly
evident from fig. 2, their centers are scattered inside the
shell namely they change in a discontinuous way on moving 
from one eigenfunction to the next. In such a situation the derivate
$\partial lnZ^s_E
\over {\partial E}$ does not even exist and the equilibrium
distribution will depend on initial excitation. Notice that for
classically chaotic conservative systems like billiards, the situation is
even worse. Indeed in such cases it is clearly seen\cite{cp} that
both the perturbed and unperturbed eigenfunctions  have a 
sparse structure and this makes even more evident the impossibility to 
derive quantum equilibrium distributions.  
The same conclusions
will follow when localization is produced by the lack of classical
ergodicity: 
namely, by the existence of
islands of stability in classical phase space (in this case too, 
temperature cannot be defined).

We would like to stress that the shape of the LDOS as well as
its width is not relevant for the problem discussed here. 
As a matter of fact, for general Hamiltonian systems of  type (1), the 
LDOS is not given by eq.(3.5) and the function 
$W_E(E_0)$ depends on the perturbation $V$.
This perturbation can be very small so that the total energy
$H=E$ can be very close to the unperturbed energy $H_O=E_0$.
In spite of this fact,  in case of ergodicity  the orbit of
system (1) will move on the whole $2N-1$ dimensional energy
surface while  the unperturbed integrable motion will take place on the
N-dimensional torus.  Analogously, in the quantum
case, what is relevant is whether or not the perturbation actually couples 
the maximal numbers of unperturbed states: namely, whether or not the
eigenfunctions are extended over the whole  energy shell.
Therefore, the relevant parameter is $\lambda$; for $\lambda>1$ one expects Fermi-Dirac (or
Bose-Einstein) statistics to hold. Instead, if $\lambda<1$, one
expects deviations from equilibrium statistical distributions
and strong dependence on initial conditions.
  
Another question concerns the number of degrees of freedom. One lesson we 
have learnt from the study of dynamical chaos 
is that the so-called thermodynamical limit is not necessary: statistical 
laws can appear in systems with very few degrees of freedom, provided they 
are ergodic and mixing. We argue that quantum statistical distributions 
will take place even in systems with few degrees of freedom, provided 
dynamical localization is absent, 
namely $\lambda>1$. Finally we would like to remark that for classically 
ergodic systems the ergodicity parameter $\lambda$ always becomes larger 
than one in the semiclassical limit, as required by the correspondence 
principle.

\section{\bf Conclusions}
 
In this paper we have attempted to elucidate the problem of the
foundations of quantum equilibrium statistical distributions at
the light of recent progress in nonlinear dynamics and in
quantum chaos.In particular we have shown that in presence of quantum
dynamical localization the equilibrium state depends on the initial
excitation and does not obey the standard quantum statistical
distributions. The corresponding problem
of energy equipartition in classical mechanics has been, at
the very start of this century, at the root of the transition to the 
quantum theory. More recently, the discovery of FPU
and KAM, that below a certain threshold the ergodic hypothesis is
not valid and therefore energy equipartition among different
degrees of freedom does not follow, has made a significant impact
in our understanding of physical systems. It is possible
that a critical re-examination of the analogous problem in
quantum mechanics, in order to justify quantum equilibrium distributions, 
may lead to interesting developments for our
understanding of the microscopic world.


\begin{figure}
\centerline{\epsfxsize=15cm \epsfbox{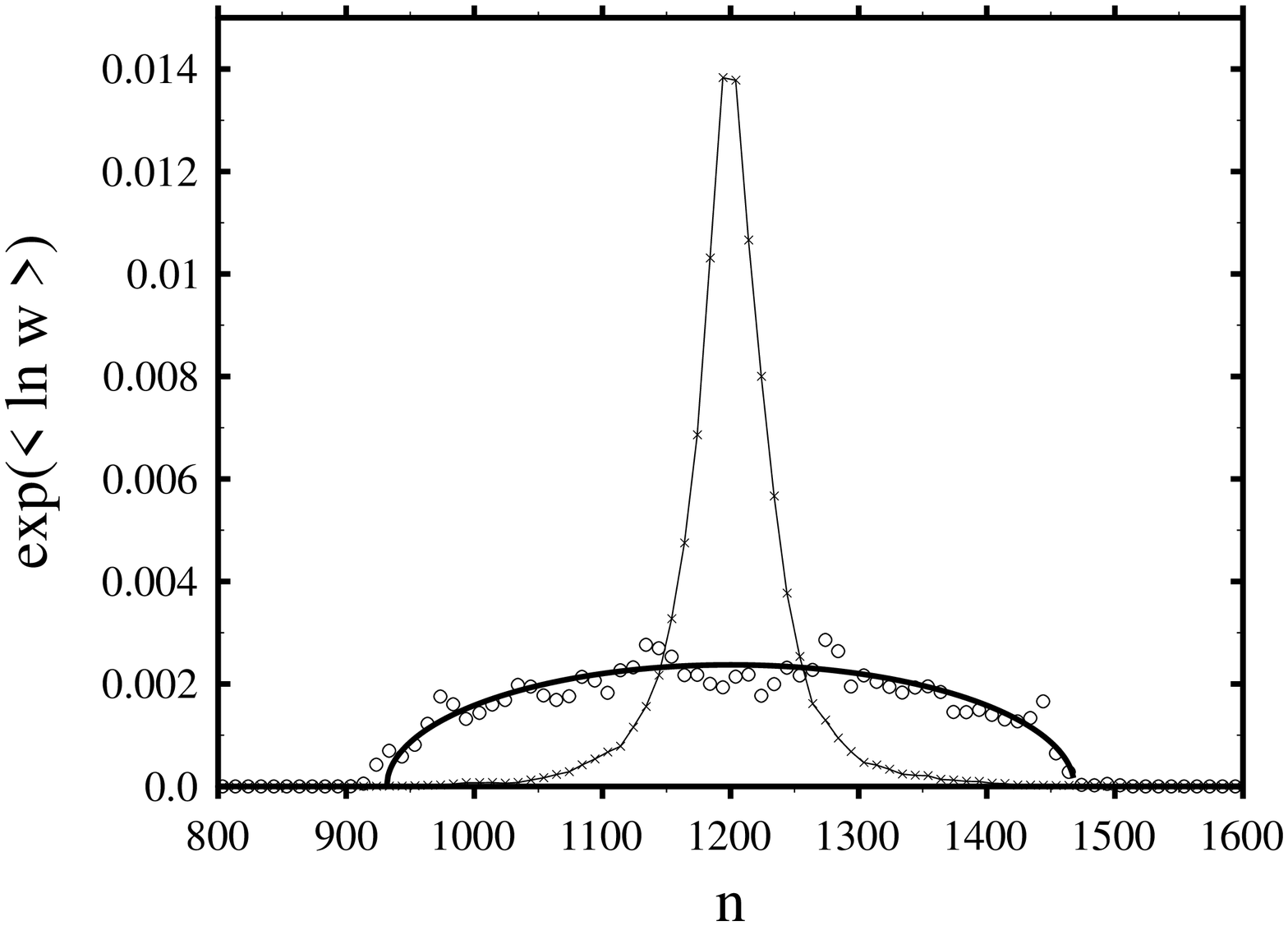}}
\vspace{4mm}                                       
\caption{Structure of a localized   
eigenfunction for a single matrix 
with parameters  
$N=2400$, $v=0.1$, $b=10$, 
$\rho=300$, $q=90$. The fat full line is the semicircle law (3.5). The  
solid line was obtained by averaging $300$ eigenfunctions with respect
to  their centers; circles, by averaging the same eigenfunctions with respect 
to the centers of their energy shells.
Here the parameter
$\lambda=0.24$, and  the average with respect to centers 
$n_c(m)$ of the distributions $W_m(n)$ shows  
a clear localization with $\beta=0.24$, 
while the other average(circles) remains close  
to semicircle, with $\beta=0.99$.} 
\end{figure}

\begin{figure}
\centerline{\epsfxsize=15cm \epsfbox{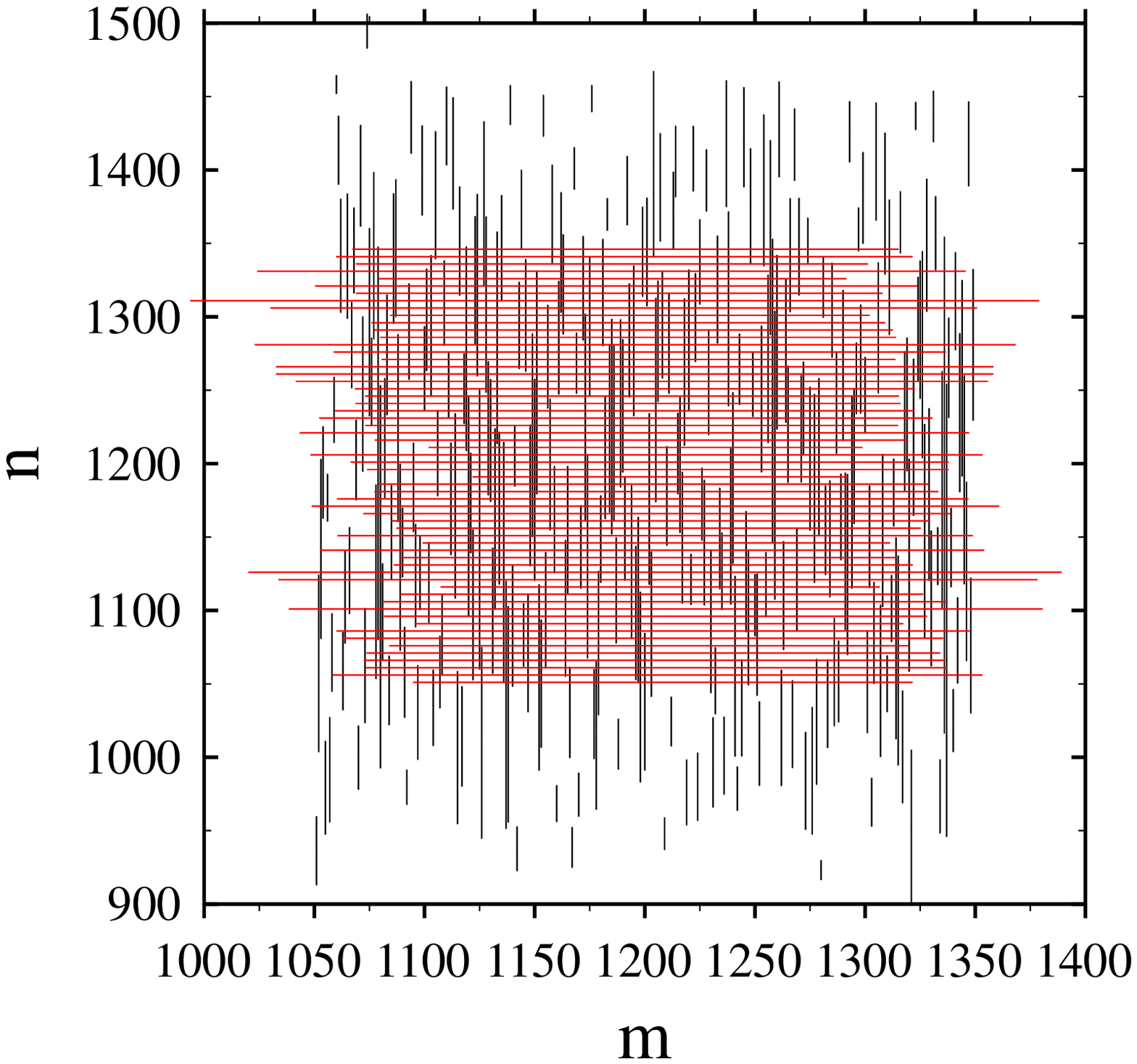}}
\vspace{4mm}                                       
\caption{A comparison of the structure of eigenfunctions and of  
LDOS in the localized case of fig.1. Solid vertical bars represent the  
widths $\Delta n$ of individual eigenfunctions over the unperturbed basis.   
Horizontal dotted lines show the size $\Delta m$ of the local spectrum for  
individual basis states. 
Although all basis states have comparable sizes,  
close to the size of the energy shell, they are very sparse ($\beta=0.20$),  
due to the fact that   
EF's are strongly localized, and irregularly scattered inside the energy shell. 
} 
\end{figure}


\begin{thebibliography}{99}
\bibitem{nota} We should note that a number of papers
have been devoted to the properties of eigenfunctions for
classically ergodic sytems. Among these we mention for example
Shnirelman\cite{SH} and Berry\cite{Be}.
\end{thebibliography}
\end{document}